\documentclass[aps,pra,showpacs,twocolumn,floatfix,a4paper]{revtex4}
\newcommand{\id}{{\sf 1 \hspace{-0.3ex} \rule{0.1ex}{1.52ex}\rule[-.01ex]{0.3ex}{0.1ex}}}

\usepackage{amsmath,amsfonts,amssymb,graphics,graphicx,epsfig,color,times,bbm}
\usepackage{theorem}
\newtheorem{definition}{Definition}

\newtheorem{theorem}{Theorem}

\begin{document}
\title{The Quantum Blackwell Theorem and Minimum Error State Discrimination}
\author{Anthony Chefles}
\affiliation{Quantum Information Processing Group, Hewlett-Packard
Laboratories, Filton Road, Stoke Gifford, Bristol BS34 8QZ, U.K.}
\email{anthony.chefles@btinternet.com}
\begin{abstract}
\vspace{0.5cm}A quantum analogue of the famous Blackwell Theorem in classical statistics has recently been proposed.  Given two quantum channels ${\cal A}$ and ${\cal B}$, a set of payoff functions have been proven to have values for ${\cal B}$ at least as high as they are for ${\cal A}$ if and only if there exists a quantum garbling channel ${\cal E}$ such that ${\cal A}={\cal E}{\cal B}$.  When such a channel ${\cal E}$ exists, we can globally compare ${\cal A}$ and ${\cal B}$ in terms of their `noisiness'.   We show that this method of channel noise comparison is equivalent to one obtained by considering the degradation of the distinguishability of states.  Here, the channel ${\cal A}$ is said to be at least as noisy as the channel ${\cal B}$ if any ensemble of states, fed into each channel and possibly entangled with ancillae, emerges no more distinguishable from ${\cal A}$ than it does from channel ${\cal B}$, where distinguishability is quantified by the minimum error discrimination probability.  We also provide a novel application to eavesdropper detection in quantum cryptography.
\end{abstract}
\maketitle
The understanding of quantum noise is of paramount importance in all areas of applied quantum information science.  For example, the security of quantum cryptosystems depends directly on the noise properties of the quantum channels used.  Also, quantum noise has a deleterious effect on quantum computations, giving rise to the phenomenon of decoherence, which is the main current obstacle to the construction of a large scale quantum computer.

One of the principal difficulties in understanding quantum noise is that, for a $D$ dimensional quantum system, a quantum channel is described by $D^{4}$ independent parameters subject to $D^{2}$ global constraints.  There is no single parameter which can be used to quantify how `noisy' a given quantum channel is.  As such, given two quantum channels, the question as to how we might determine which one is the least noisy is highly non-trivial.  Fortunately, a similar situation arises in the theory of classical channels where is the issue has been partially resolved in the following way: a channel ${\cal A }$ is at least as noisy as a channel ${\cal B}$ if ${\cal B}$ can be post-concatenated with some channel ${\cal E}$ to give ${\cal A}$, i.e. ${\cal A}={\cal E}{\cal B}$.  The channel ${\cal E}$ is referred to as a garbling channel.  A famous result in classical statistics, known as Blackwell's Theorem \cite{Blackwell}, has the following content.  It provides a set of payoff functions whose values are all at least as high for ${\cal B}$ as they are for ${\cal A}$ if and only if such a garbling channel ${\cal E}$ exists.  In what follows, we shall refer to this non-decreasing property of the payoff functions as monotonicity.

In a recent, remarkable work, Shmaya \cite{Shmaya} has obtained a quantum analogue of Blackwell's Theorem, where the channels are quantum.  Shmaya's Theorem provides necessary and sufficient conditions, in terms of the monotonicity of certain quantum payoff functions, for the existence of a quantum garbling channel ${\cal E}$ such that ${\cal A}={\cal E}{\cal B}$ for a given pair of quantum channels ${\cal A}$ and ${\cal B}$.  A quantum channel ${\cal A}$ of the form ${\cal E}{\cal B}$ for some quantum channels ${\cal E}$ and ${\cal B}$ is said, using the terminology of Wolf and Cirac, to be divisible.  These authors have investigated the properties of divisible channels in detail \cite{WC} (see also Holevo \cite{Holevo} and Denisov \cite{Denisov}.)  An important special class of divisible channels are degradable channels.  These have attracted considerable attention recently, see e.g. \cite{DW,CRS}.

In Blackwell's Theorem, the payoffs are expressed in terms of game-theoretic utility functions and thus have a direct operational interpretation which expresses the superiority of one channel over another.  While Shmaya also provides a game-theoretic setting in which his payoff functions have an operational interpretation, insofar as they are measureable, they are not of obvious general practical significance and it is not clear that they can be understood in terms of more established notions of quantum noise.  The purpose of this Letter is to describe precisely how they relate to one of the most prevalent and intuitively meaningful aspects of quantum noise: the property of decreasing the distinguishability of quantum states.  From a practical point of view, we may characterise one quantum channel, ${\cal A}$, as being at least as noisy as another quantum channel ${\cal B}$, if any ensemble of states, when fed into channel ${\cal A}$, would emerge no more distinguishable than would be the case if they were fed into channel ${\cal B}$.

Clearly, to make this idea precise, we require a measure of distinguishability, so let us choose the simplest one which can be applied to all ensembles of states: the maximum discrimination probability using the minimum error discrimination strategy \cite{Helstrom}.  We shall actually refine this idea somewhat, and allow for ensembles of states which include an external ancilla which is not fed into the channel, although the entire states are subject to the discrimination measurement.  Our main result is that the monotonicity of the maximum discrimination probabilities is equivalent to the monotonicity of Shmaya's payoff functions and thus to the existence of a garbling channel ${\cal E}$.  We then explore the practical implications of this finding, where we find a novel application to eavesdropper detection in quantum cryptography.

We begin by reviewing the aspects of Shmaya's Theorem that we shall need.  Consider four quantum subsystems, ${\alpha}$, ${\beta}$, ${\gamma}$ and ${\delta}$ with corresponding Hilbert spaces $H_{\alpha}, H_{\beta}, H_{\gamma}$ and $H_{\delta}$.  For simplicity, we take these four spaces to have equal, finite-dimensionality $D$ and shall thus frequently denote them by the generic $H$.  We denote by $B(H)$ the set of linear operators on $H$ and note that all such operators on finite-dimensional Hilbert spaces are bounded.  We also denote by $M(H)$ the set of linear maps from $B(H){\rightarrow}B(H)$. The subsystems ${\alpha}$ and ${\beta}$ constitute an environment whose joint state is described by the density operator ${\rho}$.  The subsystems ${\gamma}$ and ${\delta}$ form the main system of interest, whose state is initially described by one of two density operators ${\Psi}$ and ${\Phi}$.

\begin{figure}
\begin{center}
\epsfxsize8cm \centerline{\epsfbox{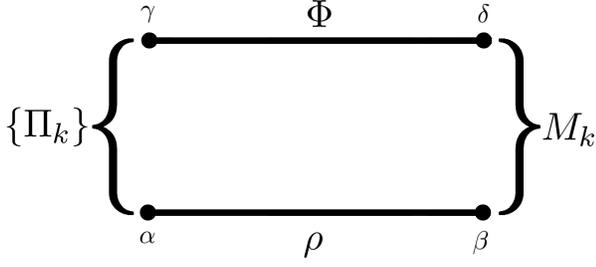}}
\end{center}
\caption{Illustration of the setup considered in Shmaya's Theorem.  The compound system ${\gamma}{\delta}$ is prepared in the joint state ${\Phi}$ while the environmental systems ${\alpha}{\beta}$ are prepared in the state ${\rho}$.  A POVM measurement $\{{\Pi}_{k}\}$ is carried out on ${\gamma}{\alpha}$ whose result, $k$, is used to choose an Hermitian operator $M_{k}$ which is then measured on ${\delta}{\beta}$.  The average, over $k$, of the expectation value of $M_{k}$ defines the payoff.  The same considerations apply to the state ${\Psi}$ of ${\gamma}{\delta}$.}
\label{figure1}
\end{figure}

We now consider playing a game, which is illustrated in Figure (\ref{figure1}). The rules are described in relation to the state ${\Phi}$ although they apply equally well to the state ${\Psi}$: \\

\noindent{\em (1)}: ${\gamma}{\delta}$ is prepared in the state ${\Phi}$ and
the environment ${\alpha}{\beta}$ is
prepared in the state ${\rho}$.\\

\noindent{\em (2)}: A measurement is performed on ${\gamma}{\alpha}$.  This will
be described by a POVM, i.e. a set of unity-resolving, positive operators ${\Pi}_{k}$
acting on $H_{\gamma}{\otimes}H_{\alpha}$, where $k{\in}\{1,{\ldots},K\}$, for some integer $K>0$.\\

\noindent{\em (3)}:  Using the result of this measurement, i.e.,
$k$, a corresponding Hermitian operator $M_{k}$ is measured on ${\delta}{\beta}$.\\

Let us now consider the expected payoff.  This is defined to be the average of the expectation values of the $M_{k}$.  Denoting this payoff by $R({\Phi},{\rho},\{M_{k}\},\{{\Pi}_{k}\})$, one finds that
\begin{equation}
R({\Phi},{\rho},\{M_{k}\},\{{\Pi}_{k}\})=\mathrm{Tr}\biggl[({\Phi}_{{\gamma}{\delta}}{\otimes}{\rho}_{{\alpha}{\beta}})\sum_{k=1}^{K}({\Pi}_{k{\gamma}{\alpha}}{\otimes}M_{k{\delta}{\beta}})
\biggr].
\end{equation}
The maximum payoff, which is to say the payoff maximised with respect to the measurement on ${\gamma}{\alpha}$, will be of particular interest to us.  It is given by
\begin{eqnarray}
&&R_{max}({\Phi},{\rho},\{M_{k}\}) \nonumber \\
&=&\max_{\{{\Pi}_{k}\}}\mathrm{Tr}\biggl[({\Phi}_{{\gamma}{\delta}}{\otimes}{\rho}_{{\alpha}{\beta}})\sum_{k=1}^{K}({\Pi}_{k{\gamma}{\alpha}}{\otimes}M_{k{\delta}{\beta}})\biggr].
\end{eqnarray}
These quantities, for all particular environment states ${\rho}$ and Hermitian operator sets $\{M_{k}\}$, are the quantum payoff functions that occur in Shmaya's quantum analogue of Blackwell's Theorem. Prior to describing this, we require the following:
\begin{definition}
${\Phi}$ is at least as good as ${\Psi}$ if
\begin{equation}
\label{Rineq}
R_{max}({\Phi},{\rho},\{M_{k}\}){\geq}R_{max}({\Psi},{\rho},\{M_{k}\})
\end{equation}
for every set of $K$ Hermitian operators $\{M_{k}\}$ on $H_{\delta}{\otimes}H_{\beta}$, for all integers $K>0$
and every density operator ${\rho}$ on $H_{\alpha}{\otimes}H_{\beta}$.  We denote this relationship by
${\Phi}{\supseteq}{\Psi}$.
\end{definition}
We are now in a position to state:
\begin{theorem}{\bf(Shmaya's Theorem)}
For any two bipartite states ${\Phi}$ and ${\Psi}$ in $B(H_{\gamma}{\otimes}H_{\delta})$,
\begin{equation}
\label{shmaya1}
{\Phi}{\supseteq}{\Psi}{\Leftrightarrow}{\Psi}_{{\gamma}{\delta}}=({\cal E}_{\gamma}{\otimes}{\id}_{\delta})({\Phi}_{{\gamma}{\delta}})
\end{equation}
\end{theorem}
for some quantum channel ${\cal E}{\in}M(H_{\gamma})$.

The analogy with Blackwell's Theorem is made apparent if we choose
\begin{eqnarray}
\label{specialPsi}
{\Psi}_{{\gamma}{\delta}}&=&({\cal A}_{\gamma}{\otimes}{\id}_{\delta})({\chi}_{{\gamma}{\delta}}), \\
\label{specialPhi}
{\Phi}_{{\gamma}{\delta}}&=&({\cal B}_{\gamma}{\otimes}{\id}_{\delta})({\chi}_{{\gamma}{\delta}}),
\end{eqnarray}
where ${\chi}$ is a faithful state \cite{DP,altepeter}.  For such states, imprinting of a quantum channel is complete and logically reversible.  For such states, we therefore obtain
\begin{equation}
\label{shmaya2}
{\Phi}{\supseteq}{\Psi}{\Leftrightarrow}{\cal A}={\cal E}{\cal B},
\end{equation}
which is the main requirement of a quantum analogue of Blackwell's Theorem.

We will shortly describe the connection between Shmaya's Theorem and quantum state discrimination.  Prior to doing so, we make the following crucial observation about Shmaya's analysis.  In proving the forward implication in (\ref{shmaya1}), Shmaya was able to make the assumption that the environment state ${\rho}$ is of a specific form.  Consider the state
\begin{equation}
\label{kdef}
|I_{D}{\rangle}=\frac{1}{\sqrt{D}}\sum_{i=1}^{D}|x_{i}{\rangle}{\otimes}|x_{i}{\rangle}
\end{equation}
where $\{|x_{i}{\rangle}\}$, with $i{\in}\{1,{\ldots},D\}$, is an
arbitrary orthonormal basis for $H_{\delta}=H_{\gamma}$.  This is a faithful state.  Then for the states ${\Psi}$ and ${\Phi}$ in Eqs. (\ref{specialPsi}) and (\ref{specialPhi}), Shmaya's Theorem admits the following strengthened definition of `as least as good as':
\begin{definition}
${\Phi}$ is at least as good as ${\Psi}$ if
\begin{equation}
\label{Rineq2}
R_{max}({\Phi},{\rho},\{M_{k}\}){\geq}R_{max}({\Psi},{\rho},\{M_{k}\}),
\end{equation}
for every set of $K$ Hermitian operators $\{M_{k}\}$ on $H_{\delta}{\otimes}H_{\beta}$ and for all integers $K>0$
where ${\chi}={\rho}=|I_{D}{\rangle}{\langle}I_{D}|$.
\end{definition}
For our purposes, it will be necessary to determine the explicit forms of the maximum payoffs for these states. We find that
\begin{eqnarray}
&&R_{max}({\Psi},{\rho},\{M_{k}\}) \nonumber \\
&=&\max_{\{{\Pi}_{k}\}}\frac{1}{D^{2}}\sum_{k=1}^{K}\mathrm{Tr}\biggl[{\Pi}_{k{\delta}{\beta}}({\cal A}_{\delta}{\otimes}{\id}_{\beta})(M^{T}_{k{\delta}{\beta}})\biggr],
\label{rmaxpsi}
\end{eqnarray}
\begin{eqnarray}
\label{}
&&R_{max}({\Phi},{\rho},\{M_{k}\}) \nonumber \\ &=&\max_{\{{\Pi}_{k}\}}\frac{1}{D^{2}}\sum_{k=1}^{K}\mathrm{Tr}\biggl[{\Pi}_{k{\delta}{\beta}}({\cal B}_{\delta}{\otimes}{\id}_{\beta})(M^{T}_{k{\delta}{\beta}})\biggr].
\label{rmaxphi}
\end{eqnarray}
Here, $T$ denotes transposition in the product basis $|x_{i}{\rangle}{\otimes}|x_{j}{\rangle}$.  It is important to note here that the set of possible sets $\{M^{T}_{k}\}$ is identical to the set of possible sets $\{M_{k}\}$, i.e. they are both the set of possible sets of $K$ Hermitian operators on $H^{{\otimes}2}$.  The forms of the expressions in Eqs. (\ref{rmaxpsi}) and (\ref{rmaxphi}) will be key to establishing our relationship between Shmaya's Theorem and minimum error state discrimination, which we shall now do.

In minimum error state discrimination, we consider a quantum system whose state is given by one of $K$ possible density operators
$\varrho_{k}$ which have a priori probabilities $P_{k}$, which is to say an ensemble of quantum states $\{\varrho_{k}, P_{k}\}$.
We consider a measurement with $K$ outcomes. The $k$th outcome,
which has corresponding POVM element ${\Pi}_{k}$, is associated
with the $k$th state, ${\varrho}_{k}$.  If we obtain the result
$k$, then we take this to signify that the initial state was
${\varrho}_{k}$.  If this is true, then our result is correct, if
not, then we have an error. The total probability of obtaining a correct result is
\begin{equation}
\label{pdisc1}
P(\{\varrho_{k},P_{k}\},\{{\Pi}_{k}\})=\sum_{k=1}^{K}P_{k}\mathrm{Tr}(\Pi_{k}\varrho_{k}),
\end{equation}
and so we see that the maximum probability of correct discrimination as
\begin{equation}
\label{pdisc2}
P_{max}(\{\varrho_{k},P_{k}\})=\max_{\{{\Pi}_{k}\}}\sum_{k=1}^{K}P_{k}\mathrm{Tr}(\Pi_{k}\varrho_{k}).
\end{equation}
The minimum error probability is $1-P_{max}(\{\varrho_{k},P_{k}\})$.  It is for this reason that the state discrimination strategy we have described is known as minimum error state discrimination \cite{Helstrom}.
\begin{figure}
\begin{center}
\epsfxsize8cm \centerline{\epsfbox{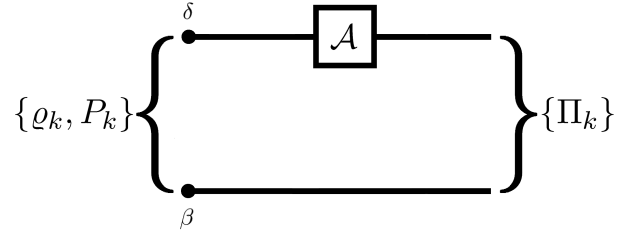}}
\end{center}
\caption{Channel noise comparison in terms of quantum state discrimination.  The ensemble of states $\{\varrho_{k}, P_{k}\}$ is processed by feeding the ${\delta}$ system into a noisy channel, here ${\cal A}$.  The resulting total output states are then discriminated using the POVM which attains the minimum error probability.  If, for every ensemble, the correct discrimination probability using the channel ${\cal A}$ never exceeds that for some other channel ${\cal B}$, then we can say that ${\cal A}$ is at least as noisy as ${\cal B}$.} \label{figure2}
\end{figure}
We may view $P_{max}(\{\varrho_{k},P_{k}\})$ as a measure of the distinguishability of the ensemble $\{\varrho_{k},P_{k}\}$.  So let us now consider distinguishability degradation as a means of comparing the `noisiness' of quantum channels.  Rather than limit ourselves to the scenario where the systems carrying the states to be distinguished have been fed entirely into the noisy channel, we allow for the possibility that the entire systems comprise also ancillae.  To be precise, consider, for example, the channel ${\cal A}$, which we take here to act on the subsystem ${\delta}$.  The entire system, comprised of subsystems ${\delta}$ and ${\beta}$, is prepared in the compound state ${\varrho}_{k}$ with a priori probability $P_{k}$.  The ${\delta}$ subsystem is then fed into channel ${\cal A}$, following which we aim to discriminate among the possible global output states $({\cal A}_{\delta}{\otimes}{\id}_{\beta})({\varrho}_{k{\delta}{\beta}})$.  This involves performing a global measurement with associated POVM elements ${\Pi}_{k}$ acting on the Hilbert space of ${\delta}{\beta}$.  This scenario is illustrated in Figure (\ref{figure2}). The maximum probability of correct discrimination is then
\begin{equation}
P_{max}(\{\varrho_{k},P_{k}\},{\cal A})=\max_{\{{\Pi}_{k}\}}\sum_{k=1}^{K}P_{k}\mathrm{Tr}\biggl[{\Pi}_{k{\delta}{\beta}}({\cal A}_{\delta}{\otimes}{\id}_{\beta})(\varrho_{k{\delta}{\beta}})\biggr].
\label{pdisc3}
\end{equation}
A natural way of comparing the noisiness of two quantum channels ${\cal A}$ and ${\cal B}$ in terms of state discrimination is as follows: we say that ${\cal A}$ is at least as noisy as ${\cal B}$, writing this formally as ${\cal A}{\leq}_{disc}{\cal B}$, if every ensemble of $D{\times}D$ bipartite quantum states emerges no more distinguishable, in terms of the maximum discrimination probability, if the ${\delta}$ system is fed into channel ${\cal A}$ than if it is fed into channel ${\cal B}$, that is
\begin{equation}
\label{pdisc4}
P_{max}(\{\varrho_{k},P_{k}\},{\cal B}){\geq}P_{max}(\{\varrho_{k},P_{k}\},{\cal A}),
\end{equation}
for all ensembles $\{{\varrho}_{k},P_{k}\}$ and all integers $K>0$.  We are now in a position to state and prove our main result
\begin{theorem}
Let ${\cal A}$ and ${\cal B}$ be two quantum channels in $M(H_{\delta})$.  Then the following statements are equivalent:\\

\noindent $(i)$  There exists a quantum garbling channel ${\cal E}{\in}M(H_{\delta})$ such that ${\cal A}={\cal E}{\cal B}$. \\
\noindent $(ii)$  ${\cal A}{\leq}_{disc}{\cal B}$.
\end{theorem}
\noindent{\bf Proof} The proof of $(i){\Rightarrow}(ii)$ is straightforward.  By assumption, ${\cal A}={\cal EB}$ and so Shmaya's Theorem implies (\ref{Rineq2}) for the the payoff functions in Eqs. (\ref{rmaxpsi}) and (\ref{rmaxphi}) where $\{M^{T}_{k}\}$ is any set of $K$ Hermitian operators acting on $H_{\delta}{\otimes}H_{\beta}$.  So consider the choice $M^{T}_{k}=D^{2}P_{k}{\varrho}_{k}$ where $\{{\varrho}_{k},P_{k}\}$ is an arbitrary ensemble of states in $B(H_{\delta}{\otimes}H_{\beta})$.  From the definition of the maximum discrimination probability in Eq. (\ref{pdisc3}), we see that (\ref{Rineq2}) leads to (\ref{pdisc4}), giving the forward implication as desired.

To prove that $(ii){\Rightarrow}(i)$, we make use of the fact for any set of $K$ Hermitian operators $\{M^{T}_{k}\}{\subset}B(H_{\delta}{\otimes}H_{\beta})$, we can define an ensemble $\{{\varrho}_{k},P_{k}\}$ where
\begin{equation}
\label{specialstate}
\varrho_{k}=\frac{M^{T}_{k}+({\epsilon}-{\Lambda}){\id}_{{\delta}{\beta}}}{\mathrm{Tr}(M_{k})+D^{2}({\epsilon}-{\Lambda})}
\end{equation}
and
\begin{equation}
\label{specialprob}
P_{k}=\frac{\mathrm{Tr}(M^{T}_{k})+D^{2}({\epsilon}-{\Lambda})}{\mathrm{Tr}(\sum_{k=1}^{K}M_{k})+D^{2}K({\epsilon}-{\Lambda})}.
\end{equation}
Here, we have ${\Lambda}={\min}_{k,|{\psi}{\rangle}}{\langle}{\psi}|M^{T}_{k}|{\psi}{\rangle}$, that is, the smallest, with respect to $k$, of the minimum eigenvalues of the $M^{T}_{k}$ (or equivalently of the $M_{k}$.)  This is guaranteed to be finite by the boundedness of these operators.  The real parameter ${\epsilon}$ may take any value that ensures the positivity of the ${\varrho}_{k}$.  We find that we require ${\epsilon}>\mathrm{Tr}(M_{k})/D$ for all $k{\in}\{1,{\ldots},K\}$.  Making use of Eqs. (\ref{rmaxpsi}), (\ref{rmaxphi}) and (\ref{pdisc3}), we see that the maximum discrimination probabilities for this ensemble have the form
\begin{equation}
P_{max}(\{\varrho_{k},P_{k}\},{\cal A})=\frac{D^{2}(R_{max}({\Psi},{\rho},\{M_{k}\})+{\epsilon}-{\Lambda}
)}{\mathrm{Tr}(\sum_{k=1}^{K}M_{k})+D^{2}K({\epsilon}-{\Lambda})},
\end{equation}
\begin{equation}
P_{max}(\{\varrho_{k},P_{k}\},{\cal B})=\frac{D^{2}(R_{max}({\Phi},{\rho},\{M_{k}\})+{\epsilon}-{\Lambda}
)}{\mathrm{Tr}(\sum_{k=1}^{K}M_{k})+D^{2}K({\epsilon}-{\Lambda})}.
\end{equation}
The identical denominators here are easily proven to be strictly positive.  It follows that (\ref{Rineq2}) and (\ref{pdisc4}) are equivalent and Shmaya's Theorem then automatically completes the proof.

There is a clear similarity between between the expressions for the maximum discrimination probability and the maximum payoff and it is very useful to know that this leads to the equivalence of channel noise comparison in terms of the structural ${\cal A}={\cal E}{\cal B}$ criterion which appears in Shmaya's Theorem and the practically-motivated one based on state discrimination.  Indeed, in our proof of the equivalence of the two inequalities (\ref{Rineq2}) and (\ref{pdisc4}), one can observe that the two aspects of the affine transformations of the underlying operators, namely scaling and shifting, correspond to the two differences between a general Hermitian operator and a density operator, these being normalisation and positivity.  This is no accident, as apart from a further scaling, the only difference between the maximum payoffs and the maximum discrimination probabilities is that the former involve the general Hermitian operators $M^{T}_{k}$ while the latter involve the positive operators $P_{k}{\varrho}_{k}$ subject to normalisation conditions on the $P_{k}$ and ${\varrho}_{k}$.

From a practical point of view, perhaps the most important consequence of our theorem is as follows.  If ${\cal A}={\cal E}{\cal B}$ for some channel ${\cal E}$, then ${\cal E}$ will never improve the quality of any information, classical or quantum, sent through it.  If, however, there is no such channel, then we have at least one specific communication task for which ${\cal A}$ is demonstrably superior.  This is minimum error discrimination for an ensemble of states for which inequality (\ref{pdisc4}) is not satisfied.

We may then consider the following application to quantum cryptography.  Suppose that Alice and Bob initially share a unidirectional quantum channel ${\cal B}$ and suspect that this arrangement has been modified by an eavesdropper, Eve, operating at the receiving (Bob's) end.  Their concern is that Eve, if she exists, is extracting information from the states that Alice sends and is transmitting to Bob modified and/or replacement states.  If so, then Eve's activities can be expressed in terms of a noisy channel ${\cal E}$ and the actual channel between Alice and Bob will be ${\cal A}={\cal E}{\cal B}$. To ascertain whether or not a garbling channel ${\cal E}$ exists, Alice and Bob must distinguish the channel ${\cal B}$ from ${\cal E}{\cal B}$.  Our theorem enables this task to be carried out for any particular ${\cal E}$, or for any distribution of garbling channels ${\cal E}$ with a known prior probability distribution (where averaging will give rise to an effective garbling channel) using minimum error discrimination among an appropriate ensemble of states.  Of course, one could instead use a single density operator which is the optimal probe state for discrimination between ${\cal B}$ and ${\cal E}{\cal B}$.  However, this state could not be used to send any information.  Our scheme enables eavesdropper detection and information transmission using the same signals.

We have described a highly significant connection between Shmaya's quantum analogue of Blackwell's Theorem and quantum state discrimination.  We have shown that two important criteria for comparison of a pair of quantum channels ${\cal A}$ and ${\cal B}$ in terms of their `noisiness' are entirely equivalent.  One is the structural criterion ${\cal A}={\cal E}{\cal B}$ for some quantum channel ${\cal E}$.  The other is the operational criterion based on the degradation of the distinguishability of states fed into these channels.  We have also investigated its practical implications.
\section*{Acknowledgements}
This work was funded by the EU project QAP.  The author thanks Tim Spiller and Bill Munro for helpful discussions and comments.

\end{document}